\documentclass[pre,twocolumn,showpacs,preprintnumbers,amsmath,amssymb]{revtex4}

\usepackage{graphicx}
\usepackage{dcolumn}
\usepackage{bm}

\begin{document}

\title{Role of vibrations in the jamming and unjamming of grains discharging from a silo}

\author{Cristian Mankoc}
\affiliation{Departamento de F\'{\i}sica, Facultad de Ciencias,
Universidad de Navarra, E-31080 Pamplona, Spain.}
\author{Angel Garcimart\'{\i}n}
\affiliation{Departamento de F\'{\i}sica, Facultad de Ciencias,
Universidad de Navarra, E-31080 Pamplona, Spain.}
\author{Iker Zuriguel}%
\affiliation{Departamento de F\'{\i}sica, Facultad de Ciencias,
Universidad de Navarra, E-31080 Pamplona, Spain.}
\author{Luis A. Pugnaloni}
\affiliation{Instituto de F\'{\i}sica de L\'{\i}quidos y Sistemas
Biol\'ogicos\\ (CONICET La Plata--UNLP), Casilla de correo 565,
1900 La Plata, Argentina.}
\author{Diego Maza}
\affiliation{Departamento de F\'{\i}sica, Facultad de Ciencias,
Universidad de Navarra, E-31080 Pamplona, Spain.}

\date{\today}

\begin{abstract}

We present experimental results of the jamming of non-cohesive particles discharged from a flat bottomed silo subjected to vertical vibration. When the exit orifice is only a few grain diameter wide, the flow can be arrested due to the formation of blocking arches. Hence, an external excitation is needed to resume the flow. The use of a continuous gentle vibration is a usual technique to ease the flow in such situations. Even though jamming is less frequent, it is still an issue in vibrated silos. There are, in principle, two possible mechanisms through which vibrations may facilitate the flow: (i) a decrease in the probability of the formation of blocking arches, and (ii) the breakage of blocking arches once they have been formed. By measuring the time intervals inside an avalanche during which no particles flow through the outlet, we are able to estimate the probability of breaking a blocking arch by vibrations. The result agrees with the prediction of a bivariate probabilistic model in which the formation of blocking arches is equally probable in vibrated and non-vibrated silos. This indicates that the second aforementioned mechanism is the main responsible for improving the flowability in gently vibrated silos.

\end{abstract}


\maketitle

\section{Introduction}
\label{sect:intro}

Granular materials present interesting and unusual physical
properties \cite{duran,RevJaeger}. As the collisions among grains
are highly dissipative, dynamical states (such as a granular flow)
need a continuous energy supply. Otherwise, dissipation quickly
stops the motion. Even with a sustained energy input, granular
flows can --in some cases-- get jammed and halt \cite{liu:98,cates}.

A classical example of a system prone to jamming is a silo discharged by gravity. The flow of particles may be suddenly arrested by the formation of a blocking arch (or vault) if the size of the outlet does not exceed a few times the size of the particles \cite{Sakaguchi, toprl, topre0, Zuriguel1, Zuriguel2, ToPRE, AlvaroEPL}. Controversy still goes on about whether there exists or not a certain size of the orifice above which jamming is not possible. It seems that such a critical size can be found for a 3D silo \cite{Zuriguel2} but not in a quasi-2D setup \cite{ToPRE, AlvaroEPL}. Whatever it may be, in many cases it is interesting or unavoidable to reduce the size of the orifice in order to limit the flow rate; and the price to pay is an increased jamming probability. Hence the significance of the methods used to improve the flowing of grains, among which vibrations (either local or global) applied to the silo are quite common \cite{Lindemann,Schmitz,Haquette,Evesque1993,Wassgren,Chen,Pacheco,janda2}. Unfortunately, clogging still occurs. During the discharge of a vibrated silo through a small orifice, blocking arches may form. These structures can stop the flow even when the silo is being vibrated. However, the flow may be spontaneously restarted thanks to the destabilization of a blocking arch induced by the continuous excitation. Eventually, some robust jam will develop, in the sense that the continuous vibration is unable to restart the flow. A stronger external perturbation is then needed to unjam the system.

Previous works \cite{Zuriguel1, Zuriguel2} in non vibrated silos characterized the jamming by measuring the amount of grains delivered between two consecutive jams. For lack of a better term to designate this quantity, it was called an \emph{avalanche}. We have borrowed this word because it conveys the idea of a granular flow that starts and stops abruptly, but should not be confused with the rapid landslide or flood of material along a slope, and in particular, on the surface of granular piles \cite{pudasaini,daerr}. In the discharge of a silo, the avalanche size is the number of grains that flows through the orifice from the moment when the outpouring starts until the formation of a blockage that arrests the flow. In the absence of vibration, the avalanche size is a well defined quantity, as the arches or vaults that block the orifice are robust: once the flow is stopped, it does not resume by itself even after a long waiting time. It has been shown that in silos of two and three dimensions the avalanche size distribution displays an exponential tail. This can be understood if each grain has a probability $p$ of passing through the outlet without forming a blocking arch that is constant during the avalanche \cite{Zuriguel1, Zuriguel2}. In a recent paper Janda \emph{et al.} \cite{AlvaroEPL} related the value of this probability in a quasi-2D silo to the arch size distribution within a 2D static granular layer.

When vibrations are applied, the notion of avalanche is further complicated due to the fact that arches blocking the orifice can break down after a certain time. This results in an intermittent flow; in this case, in order to conclude that the flow has stopped permanently, one has to wait for a long time after grains cease to come out from the silo. The avalanches can contain themselves short lapses during which the flow has temporarily stopped, due maybe to arches that blocked the exit and were broken apart by the vibrations shortly afterwards. In Sect.~\ref{sec:avasize} we will explain how we choose a meaningful waiting time for the jams.

In this work, we present experimental results on the avalanche size distribution that allows for a comparison between vibrated and nonvibrated silos. We are able to provide the jamming probability in vibrated silos, and formulate a simple model that extends previous results and allows us to quantify the effect of vibrations. It can be conjectured that the application of vibrations to a silo may lead to two mechanisms that could cause a reduction of robust jamming events: (i) a decrease in the probability of the arch formation, which are the cause of blockages, and (ii) the loss of stability of these arches after they are generated. We will show that the second mechanism is the responsible for the appearance of larger avalanches when a gentle vibration is applied.

\section{Experimental setup}

\begin{figure}
{\includegraphics[width=0.8\columnwidth]{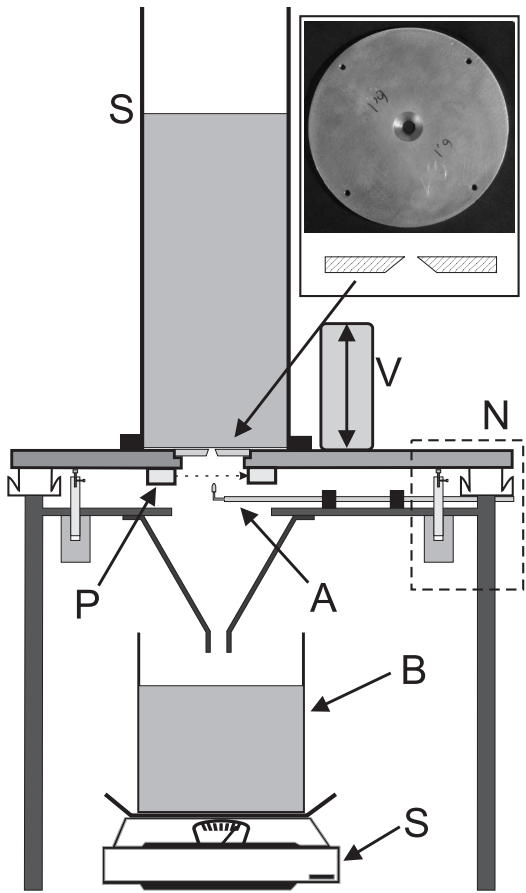}}
\caption{ \label{fig:sistexp} Experimental device. S: silo; V: pneumatic vibrator; N: damping and isolation system (air cushions and valves); P: photosensor; A: compressed air duct; B: box; S: scales. The inset shows a photograph (from the bottom) and a sketch of the cross section of the changeable part in which the hole is bored.}
\end{figure}

The experimental setup (Fig.~\ref{fig:sistexp}) comprises a flat bottomed cylindrical silo with a circular hole of diameter $D$ in its base. The part making the outlet orifice can be replaced, and in this way $D$ can be changed. The silo is filled with Delrin spheres (diameter $d = 3.00 \pm 0.01$ mm and weight $w = 18.9 \pm 0.03$ mg). The dimensionless size $R$ of the orifice is defined as $R\equiv D/d$. The diameter of the silo is more than $30$ times the particle diameter in order to prevent that the lateral walls influence the jamming probability \cite{Hirshfeld}. The silo is routinely refilled so that the height of the granular material is always more than twice the diameter of the silo. A pass-through photosensor was placed just below the outlet in order to detect whether beads are flowing or not at every time; the resolution is such that one single bead cannot go through it undetected. At the bottom of the silo, a box collects the grains that fall from the exit. This box is placed on an electronic scales, with an accuracy such that a single bead can be resolved. When the flow is arrested (see below the criterion we use to define that the flow has stopped permanently), the weight of the avalanche is obtained from the scales, and its size in number of beads is calculated. Then, a new avalanche is triggered with a jet of compressed air from beneath the orifice, a technique used in previous works \cite{Zuriguel1,Zuriguel2}.

The whole silo is continuously shaken by a pneumatic vibrator actuating in the vertical direction. In order to facilitate the vibrating movement and to isolate it from the measuring devices, the structure is supported by three air cushions, controlled through mechanical valves. We have carried out the measurements at a fixed effective dimensionless acceleration $\Gamma=a_{eff}/g=0.22$ and at a frequency of $110$ Hz. Here $a_{eff}$ is the rms value of the acceleration as measured by an accelerometer attached to the silo base, averaged over 200 cycles. In fact, this experimental set-up is quite similar to another used in previous works (see \cite{Zuriguel2}) with the addition of the pneumatic vibrator and the air cushions on which the silo rests.

\section{Avalanche size distribution and jamming probability}
\label{sec:avasize}

\begin{figure}
{\includegraphics[width=\columnwidth]{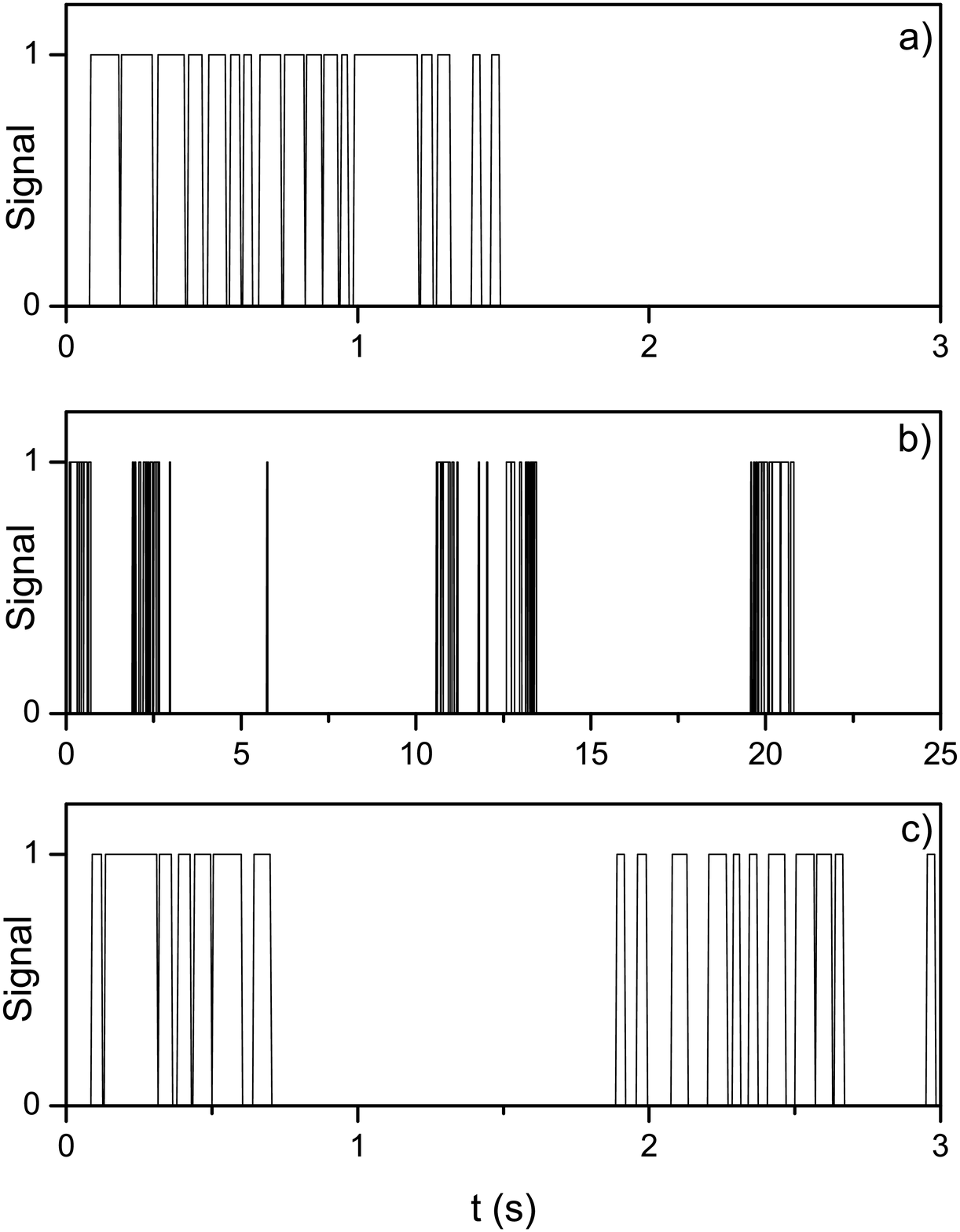}}
\caption{\label{fig:sig_photo} Signal from the photosensor at the silo exit: a value of one indicates that a particle is blocking the beam, zero means that the beam is unobstructed. \textbf{(a)} Nonvibrated silo. \textbf{(b)} Vibrated silo. \textbf{(c)} A zoom of the signal shown in (b) during the first three seconds, the same time stretch than in (a). All the data shown have been obtained for an orifice size $R=3.05$.}
\end{figure}

It is interesting to begin by inspecting a typical signal from the photosensor, which is binarized to one or zero depending on whether or not a particle is blocking the light beam. Two signals (each one registered over the duration of a single avalanche) are displayed in Fig.~\ref{fig:sig_photo}: one of them was obtained when the silo was being vibrated (Fig.~\ref{fig:sig_photo} \textbf{b}), and the other was taken with the vibration off (Fig.~\ref{fig:sig_photo} \textbf{a}). In the latter case, there are short interruptions of the flow; but once the orifice gets jammed, the flow is arrested permanently. On the other hand, when the silo is being vibrated, there are stretches during which grains are flowing separated by time intervals when the flow is interrupted. These can be significantly longer than in the nonvibrated case. This means that a jam has developed, but the vibrations are able to restart the outpouring. Eventually a blockage is formed that stops the flow permanently because the external vibration cannot break down the blocking arch; at least within our experimental time scales. In this sense, Fig.~\ref{fig:sig_photo} \textbf{b} could be described as if there were avalanches inside the avalanche. Fig.~\ref{fig:sig_photo} \textbf{c} is a zoom of Fig.~\ref{fig:sig_photo} \textbf{b} that reveals that the form of the signal of the vibrated silo, in a short interval during which the material is flowing, is similar to the one displayed by the nonvibrated silo.

As we mentioned in Sect.~\ref{sect:intro}, the avalanche size $s$ is the number of particles fallen between two consecutive jams. For the nonvibrated silo, the end of the avalanche is easily detected, as once the silo is jammed the flow does not restart by itself. In the vibrated case, we consider that the silo is jammed whenever the outflow stops for a time longer than $100$ s. We have seen that is extremely rare, for the vibrations used in our experiments, that the silo resumes its discharge after being jammed for more than this amount of time. Up to a certain point, this value is arbitrary (in the sense that one could have decided to wait for a longer time). However, the choice was made because the results presented here do not change significantly if a sufficiently long time cutoff is taken. Of course, it must be understood that this value can change a little with experimental conditions, such as $\Gamma$.

\begin{figure}
{\includegraphics[width=\columnwidth]{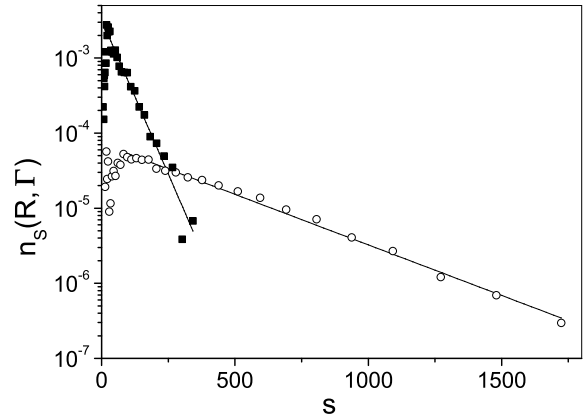}}
\caption{\label{fig:avalanche_histogram} Probability $n_s(R,\Gamma)$ of finding an avalanche of size $s$ for an orifice size $R=3.02$, without vibration ($\Gamma=0$, $\blacksquare$) and with vibration ($\Gamma=0.22$, $\circ$). The solid lines correspond to fits using Eq. (\ref{2}). For the nonvibrated silo, the fitting parameters are $p=0.981$ and $q=0$. For the vibrated silo $p=0.981$ and $q=0.836$.}
\end{figure}

For each value of $R$, the size of about $3000$ avalanches has been measured for the vibrated and the nonvibrated silo. We define $n_s(R,\Gamma)$ as the probability of finding an avalanche of $s$ grains in a vibrated silo with an opening of size $R$ and acceleration $\Gamma$. As can be seen in Fig.~\ref{fig:avalanche_histogram}, the PDF of the avalanche size displays an exponential tail for large avalanches. The comparison between the PDFs for the vibrated and nonvibrated cases (shown in the same figure for a particular value of $R$) reveals that the use of vibrations increases the size of the avalanches. However, the shape of the distribution remains exponential for large avalanches. We have found this behavior for all the outlet sizes explored ($1.7<R<4.0$). We emphasize here that jamming has not been observed for $R>5$ in the nonvibrated silo \cite{Zuriguel2}.

\begin{figure}
{\includegraphics[width=\columnwidth]{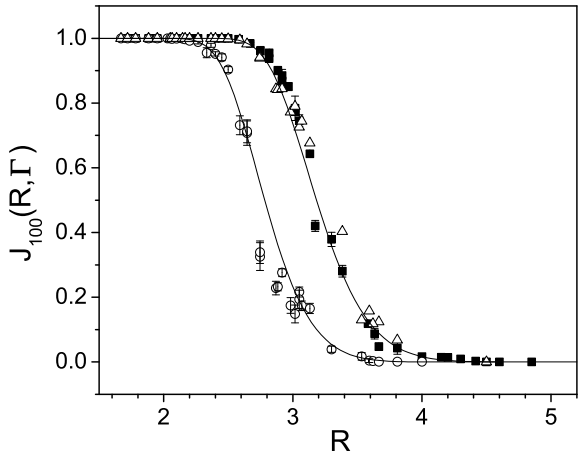}}
\caption{\label{fig:J100} Jamming probability $J_{100}(R,\Gamma)$,
\emph{i.e.} the probability that the silo gets jammed before $100$
particles flow through the outlet, as a function of $R$, for the
vibrated ($\circ$) and the nonvibrated ($\blacksquare$) silo. The
lines correspond to a fit using Eq. (\ref{J100}). The ``weak jamming'' probability $J_{100}^{\textrm{w}}(R)$ for the avalanches within an avalanche is also shown ($\triangle$).}
\end{figure}

The jamming probability $J_N(R,\Gamma)$ is defined as the probability that the silo jams before $N$ particles fall through the orifice for given $R$ and $\Gamma$. In other words, $J_N(R,\Gamma)$ is the probability of finding an avalanche smaller than $N$, \emph{i.e.}
$J_N(R,\Gamma)=\sum_{s=0}^{N}{n_s(R,\Gamma)}$ \cite{Zuriguel1}. In
Fig. \ref{fig:J100} we show the instance $J_{100}(R,\Gamma)$, obtained from the experiments with both the vibrated and the nonvibrated silo. $J_{100}(R,\Gamma=0)$ is close to one for small values of $R$ and falls to zero rather sharply in the interval $2.8 < R < 4.0$. In 3D silos it has been shown that for $N \rightarrow \infty$ the fall tends to a Heaviside function centered in $R \approx 5$ \cite{Zuriguel1,Zuriguel2}.

The vibrated case displays the same trend; however, the sharp decrease in $J_{100}(R,\Gamma=0.22)$ occurs at lower values of $R$ ($2.5<R<3.5$). It is clear that the vibration particularly reduces the jamming probability in the range of opening sizes where the transition from high to low jamming probability occurs.

For a stationary silo, the functional dependence of the mean avalanche size $\langle s\rangle$ with $R$ was shown to be compatible with a power law divergence \cite{Zuriguel2}:

\begin{equation}
\langle s\rangle= \frac{C}{(R_c-R)^\gamma} \label{divergencia}
\end{equation}

\begin{figure}
\includegraphics[width=0.9\columnwidth]{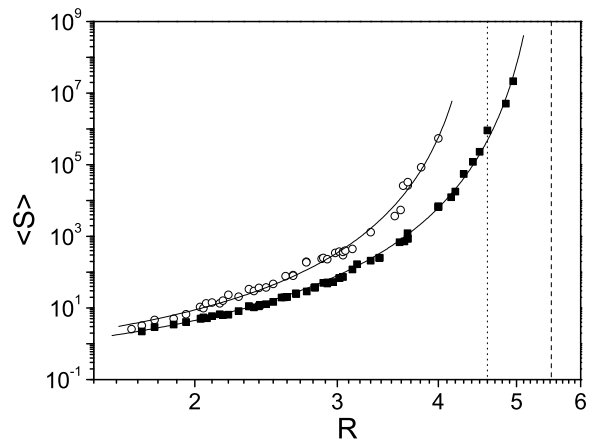}
\caption{\label{fig:mean_avalanche} Mean avalanche $\langle s\rangle$ versus the dimensionless radius $R$ for the vibrated ($\circ$) and nonvibrated ($\blacksquare$) silo. The solid lines correspond to fits using Eq. (\ref{divergencia}) with the parameters given in the text. The dotted line and the dashed line correspond to the values of $R_c$ used in the fits.}
\end{figure}

Let us assume this dependence also for vibrated silos. The experimental data can be fitted with this equation, as shown in Fig.~\ref{fig:mean_avalanche}. The values obtained for the fitting parameters are $C=2.3\pm0.7 \times 10^5$, $R_c=5.52\pm0.03$ and $\gamma=8.62\pm0.19$ for the nonvibrated case, whereas $C=1.21\pm0.3 \times 10^4$, $R_c=4.6\pm0.1$ and $\gamma=7.56\pm0.52$ for the vibrated case.

The functional form of the jamming probability $J_N(R,\Gamma)$ can be related to $\langle s\rangle$. This was done for a non vibrated silo, as explained in \cite{AlvaroEPL}. Hence, using Eq. (\ref{divergencia}), an expression of $J_N(R,\Gamma)$ can be obtained:

\begin{equation}
\displaystyle J_N(R,\Gamma)=1-e^{ -N (R_c-R)^\gamma / C}
\label{J100}
\end{equation}
which reveals an excellent agreement with the experimental data (see Fig. \ref{fig:J100}) if the values of the fitting parameters obtained from Eq. (\ref{divergencia}) are used.

\section{Probabilistic model for arch formation and breakage}

In order to describe the intermittent flow in a vibrated silo, one can assume that arches may form at the exit just as if the vibrations were absent. If the vibration breaks down the arch, the discharge resumes. Eventually, an arch strong enough can appear such that the vibration is not able to break it, at least for a considerable amount of time. As explained above, if an arch lasts for more than 100 seconds, it is assumed that it is robust enough and the flow is then said to have been stopped. Alternatively, it could also be conjectured that somehow the vibration inhibits the formation of arches. A reduced probability of arch formation would also result in longer avalanches. We now focus on answering the question of whether a gentle vibration diminishes the arch formation probability or, instead, operates by destabilizing blocking arches formed at the exit. Of course, a mix of the two mechanisms may be in place; we can still investigate if one of them dominates.

Let us extend a previous model \cite{Zuriguel1} to the case of a vibrated silo. We introduce a bivariate probability distribution to represent the behavior of a vibrated silo. We define $p(R,\Gamma)$ as the probability that a single grain passes through the outlet without forming a blocking arch along with its neighbors during continuous flow. (In the following we will drop the functional dependence of $p$ in the notation). Therefore $1-p$ accounts for the probability that a particle does get involved in the formation of an arch as it reaches the outlet. This probability $p$ is assumed to be the same for all grains and independent of discharge history.

Now, we introduce a new variable $q(R,\Gamma)$, which represents the probability that, once a blocking arch is formed, a particle flows through the exit as a consequence of the destabilization of the arch due to the continuous vibration. The probability $q$ is a measure of the ``arch breakage probability'', and does not take into account the
time that destabilization may require for a given arch. Therefore,
the probability that a blocking arch remains stable ---i.e., no more particles will ever flow out of the silo--- will be $1-q$. We assume that all arches blocking a given orifice have the same breakage probability $q$ for a given vibration intensity. This is a fair approximation since, for a particular $R$, the blocking arches are expected to have a typical size. Notice that this might not be true in hoppers, where arches form at different positions along the hopper walls, and hence a wide distribution of arch sizes may be present for a given exit size.

The probability $n_s(R,\Gamma)$ that an avalanche consists of $s$
grains can be obtained as
\begin{widetext}
\begin{align}
n_s(R,\Gamma) &= p^s(1-p)(1-q)\frac{s!}{(s-0)!0!}  \hspace{10.3em} \longleftarrow \textrm{a single blocking arch forms and remains stable.}\notag\\
&+p^{s-1}(1-p)q(1-p)(1-q)\frac{s!}{(s-1)!1!} \hspace{6em} \longleftarrow \textrm{an arch forms and breaks, then a new arch forms}\notag\\
&\hspace{23em}\textrm{  and remains stable.}\notag\\
&+p^{s-2}(1-p)q(1-p)q(1-p)(1-q)\frac{s!}{(s-2)!2!} \hspace{2.5em} \longleftarrow \textrm{two arches form and break, an arch forms and } \notag\\
&\hspace{23em}\textrm{remains stable.}\notag\\
&+...\notag\\
&+p^{s-k}[(1-p)q]^k(1-p)(1-q)\frac{s!}{(s-k)!k!} \hspace{5em} \longleftarrow \textrm{$k$ arches form and break, an arch forms and } \notag\\
&\hspace{23em}\textrm{remains stable.}\notag\\
&+[(1-p)q]^s(1-p)(1-q)\frac{s!}{0!s!} \hspace{9.5em} \longleftarrow \textrm{$s$ arches form and break, an arch forms and } \notag\\
&\hspace{23em}\textrm{remains stable.}\notag\\
\label{1}
\end{align}
\end{widetext}

In Eq. (\ref{1}) the factor $s!/((s-k)!k!$ in each term accounts for the number of permutations with repetitions of the $s-k$ events of probability $p$ (in which a grain flows through the orifice when there is no blocking arch present) and the $k$ events of probability $(1-p)q$ (in which a grain forms a blocking arch with its neighbors followed by the flow of a grain due to arch destabilization). These events can happen in any order but always add up to the desired number $s$ of grains flown. Each of these permutations will correspond to a set of $k$ jams that are unstable under the applied vibration, separated by the continuous flow of grains. For example, the last term in Eq. (\ref{1}) corresponds to the extreme case where the flow of every single particle is followed by the formation of an arch that then breaks due to the vibration. As mentioned above, the avalanche finishes when the last clogging happens and remains stable [probability $(1-p)(1-q)$]. From Eq. (\ref{1}), $n_s(R,\Gamma)$ can be written as:

\begin{align}
n_s(R,\Gamma) &= (1-p)(1-q)
\sum_{k=0}^{s}{p^{s-k}[(1-p)q]^k\frac{s!}{(s-k)!k!}} \notag\\
&= (1-p)(1-q)[p+(1-p)q]^s. \label{2}
\end{align}

It is simple to show that $p+(1-p)q<1$. Therefore, $n_s(R,\Gamma)$ is a decreasing exponential function of $s$, in excellent agreement with the experimental results shown in Fig.~\ref{fig:avalanche_histogram}. We can see that $n_s$ is properly normalized, i.e., $\sum_{s=0}^{\infty}{n_s(R)}=1$, since this sum is a geometrical series with ratio $r<1$ and it converges to one.

Thus, we can calculate the mean avalanche size, defined as

\begin{align}
<s>(R,\Gamma) &= \sum_{s=0}^{\infty}{s n_s(R)}. \label{3}
\end{align}

By inserting Eq. (\ref{2}) in Eq. (\ref{3}) we obtain a Gabriel's staircase \cite{Mathworld}. Therefore

\begin{equation}
\displaystyle <s>(R,\Gamma) = \frac{p+(1-p)q}{(1-p)(1-q)}
\label{4}
\end{equation}

\section{Estimate of the arch breakage probability}

Equations (\ref{2}) and (\ref{4}) reduce to the expression for the nonvibrated case \cite{AlvaroEPL} setting $q=0$. Equation (\ref{2}) yields an exponential form of $n_s$ for both vibrated ($q\neq0$) and nonvibrated ($q=0$) silos. In Fig.~\ref{fig:avalanche_histogram} we show the fit of the results for the nonvibrated case using Eq. (\ref{2}) with $q=0$ and $p=0.981$. For the vibrated case the data are fitted using Eq. (\ref{2}) with the same value of $p$ but $q=0.836$. This suggests that $p$ may well be independent of $\Gamma$, at least for gentle vibration, which in turn implies that the probability $1-p$ of arch formation is not affected by vibration. Hence, the enlarged sizes of the avalanches obtained for the vibrated silo would be due solely to the appearance of a probability $q$ associated to arch destabilization.

\begin{figure}
\includegraphics[width=\columnwidth]{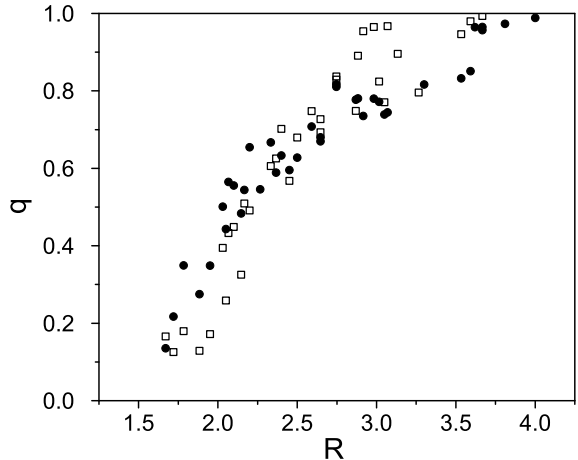}
\caption{\label{fig:zeta} Probability $q$ that a
blocking arch is destabilized due to the vertical vibration as a
function of the opening size $R$. Results obtained with the
probabilistic model, in which $p$ does not depend on $\Gamma$
(\textit{solid circles}). Results obtained from the analysis of the weak jams (\textit{open squares}). The data correspond to several sets of experiments; for some of them both methods were implemented, while for others only one of the measurements was performed.}
\end{figure}

In what follows we estimate $q$ assuming that $p$ does not depend on the external vibration. According to Eq. (\ref{4}), we can obtain $p(R,\Gamma=0)$ from the mean avalanche size of the nonvibrated silo as $\displaystyle p(R)\equiv p(R,\Gamma=0)={\langle s \rangle}_{R,\Gamma=0}/[1+{\langle s \rangle}_{R,\Gamma=0}]$. Thus, from Eq. (\ref{4}), $q$ can be calculated for $\Gamma > 0$ with the mean avalanche size for both cases as:

\begin{equation}
\displaystyle q(R,\Gamma\neq0)=\frac{{\langle
s\rangle}_{R,\Gamma\neq0} - {\langle s \rangle}_{R,\Gamma=0}}{{\langle s \rangle}_{R,\Gamma\neq0}+1}
\label{eq3}
\end{equation}

In Fig. \ref{fig:zeta} we represent with solid circles the values of $q$ obtained through Eq. (\ref{eq3}) by using the measured values of $\langle s \rangle$, for a vibrated and a nonvibrated silo. It is not surprising that $q$ increases with $R$, since big arches --needed to block big orifices-- are expected to be less robust against small perturbations than small arches. The open squares in Fig. \ref{fig:zeta} correspond to an independent estimate of $q$ measured as explained below.

\section{The internal structure of avalanches}
\label{intraavalanchas}

\begin{figure}
\includegraphics[width=\columnwidth]{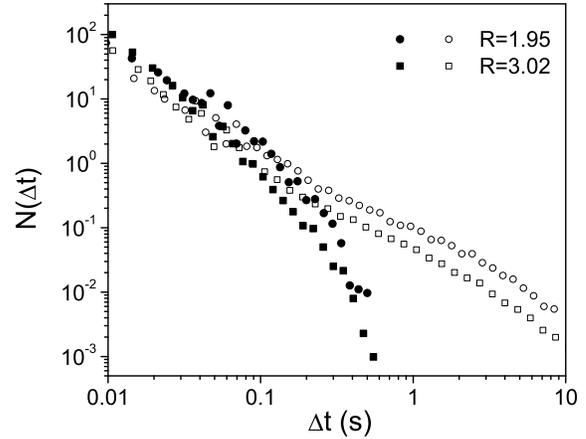}
\caption{\label{fig:unstable_jams} Normalized histograms of the
time intervals ($\Delta t$) within an avalanche during which there
are no particles flowing through the orifice. Closed (open) symbols represent the data obtained with the nonvibrated (vibrated) silo for the values of $R$ indicated in the legend. For each $R$, measurements have been carried out within, at least, 3000 avalanches.}
\end{figure}

It is possible to obtain an additional, independent estimate of the arch breaking probability, motivated by the observation of a behavior that has already been pointed out (Fig.~\ref{fig:sig_photo}): during the flow of an avalanche, short clogs are registered. We have measured the time intervals $\Delta t$ during which no beads cross the exit orifice. We represent in Fig.~\ref{fig:unstable_jams} the normalized distribution of $\Delta t$ for two different values of $R$, and these for a vibrated and a nonvibrated experiment. Let us remark that it is not judicious to consider intervals much shorter than the time it takes for a bead to traverse the orifice due to gravity (which is $t=\sqrt{\phi / g}$, about 0.02 s for the grains considered here).

The first noticeable feature in Fig.~\ref{fig:unstable_jams} is that for a nonvibrated silo, the probability of finding a time interval during which the flow is temporarily stopped decreases abruptly with $\Delta t$ (note the logarithmic scale in the plot). With vibrations, things are different. The histograms of Fig.~\ref{fig:unstable_jams} display a finite number of events for large $\Delta t$, indicating that within the avalanches longer intervals are present during which no grain comes out from the exit orifice. These events were not observed in the nonvibrated silo (nor in recent results obtained in a 2D silo \cite{fluctuacionesflujo}). These long events suggest that there are flow interruptions that might have resulted in the end of an avalanche should the silo had not been vibrated. The distribution of $\Delta t$ smaller than $0.1s$ shows similar behavior for vibrated and nonvibrated silos. Clearly, this part of the distribution corresponds to the temporal scales associated with the ``interparticle'' time intervals in the continuous flow \cite{fluctuacionesflujo}. Figure \ref{fig:unstable_jams} suggests that a temporal cutoff can be established to separate the vibrated and the nonvibrated dynamics of a few tenths of a second (where the distributions for the vibrated and nonvibrated cases begin to differ). We decided to take $\Delta t = 0.5 s$ as this cutoff. Although this choice is somehow arbitrary, we checked that the results presented below do not depend strongly on this value (for example, $\Delta t=0.25 s$ and $\Delta t=1 s$ yield similar outcomes).

Following this line of reasoning, let us call a \emph{weak jam} all the events corresponding to $\Delta t>0.5 s$. Then we can readily count the number of weak jams ($j_{\textrm{w}}$) and take that value as an indication of the number of times an arch has been formed and broken due to vibration. Since the number of robust jams ($j_{\textrm{r}}$) corresponds to the total number of avalanches studied for given $R$ and $\Gamma$, $q(R,\Gamma)$ can be obtained by dividing the number of weak jams $j_{\textrm{w}}$ by the number of total jams (weak plus robust): $\displaystyle q(R,\Gamma)=j_{\textrm{w}}/(j_{\textrm{w}}+j_{\textrm{r}})$.

In Fig. \ref{fig:zeta} the results for $q(R,\Gamma=0.22)$ obtained with this direct measure are reported (open squares). The agreement with the values of $q$ obtained from the probabilistic model under the assumption of the independence of $p$ from $\Gamma$ is reasonable. We then claim that the probability $1-p$ that a particle forms a blocking arch along its neighbors at the time of passing through the exit during the continuous flow does not depend on the external vibrations applied to the system, at least for the gentle vibration intensity we used. Therefore, the main role of vibration is to break blocking arches once they are formed.

It is now possible to measure the size of the avalanches inside the avalanches, with which we refer to the amount of material unloaded in the vibrated silo between two weak jams, as explained above. From these results we obtain the avalanche size distribution inside an avalanche, and the corresponding weak jamming probability $J_N^{w}(R,\Gamma$). Since we assumed that the cutoff chosen to identify weak jams is such that these truly correspond to the jams that would be stable if vibrations were absent, we expect the probability distribution of those avalanches in a vibrated silo to display the same properties as normal avalanches in a nonvibrated silo. In Fig. \ref{fig:J100} we can see this is indeed the case, since $J^{\textrm{w}}_{100}(R,\Gamma=0.22)$ ($\triangle$) coincides with $J_{100}(R,\Gamma=0)$ ($\blacksquare$).

\section{Conclusions}

We have reported experimental results of the avalanche size distribution and jamming probability in the discharge of a vertically vibrated silo. The first result that becomes apparent is that the presence of vibrations significantly increases the size of the avalanches. Additionally, we have shown that in a vibrated silo the discharge of grains is intermittent, due to the appearance of temporary blockages of the exit orifice. Taking as a starting point a previous model for a nonvibrated silo, we have extended it and explained the new results by introducing a bivariate probability distribution taking into account (a) the probability $p$ that a particle pass through the outlet without forming an arch with its neighbors during continuous flow, and (b) the probability $q$ that a particle flows through the exit --once and arch has been formed-- thanks to the destabilization promoted by vibration. Assuming that $p$ does not depend on the presence of vibrations, which is a fair assumption for gentle vibrations, the values of the mean avalanche size for vibrated and nonvibrated silos provide a straightforward estimation of $q$ for different values of $R$. Besides, a direct measure of $q$ has been carried out by recording the time interval within each avalanche during which grains do not flow through the exit. The two measurements agree remarkably well, indicating that the main effect of vibration is the breakage of arches once they are formed without affecting significantly the arch formation probability.

Additionally, the result of the monotonic increase of $q$ with respect to the outlet size displays an expected behavior: the greater the arch, the more unstable against small perturbations. Further measurements of the stability of arches must be carried out for different vibration frequencies and strong amplitudes to asses the full range of vibration parameters for which this explanation is valid, \emph{i.e.}, a rigorous definition of what a gentle vibration means must be provided.

\begin{acknowledgments}
This work has been financially supported by Projects FIS2008-06034-C02-01 (Spanish Government), A/9903/07 (AECI) and PIUNA (Universidad de Navarra). C. M. thanks Asociaci\'on de Amigos de la Universidad de Navarra for a scholarship. L. A. P. acknowledges financial support from CONICET (Argentina).
\end{acknowledgments}

\end{document}